\newcommand{\CLASSINPUTinnersidemargin}{1in}
\newcommand{\CLASSINPUToutersidemargin}{1in}
\newcommand{\CLASSINPUTtoptextmargin}{1in}
\newcommand{\CLASSINPUTbottomtextmargin}{1in}
\documentclass[10pt,journal,comsoc]{IEEEtran}
\usepackage[T1]{fontenc}
\usepackage[utf8]{inputenc}
\usepackage{amsmath,mathtools}
\usepackage[cmintegrals]{newtxmath}
\usepackage{graphicx}
\usepackage{booktabs,array,microtype,needspace}
\usepackage{cite}
\usepackage{xurl,hyperref}
\hypersetup{hidelinks,pdftitle={A Particle-Based Multiuser Communication Network with a Budget Limit}}
\newcolumntype{L}[1]{>{\raggedright\arraybackslash}p{#1}}
\allowdisplaybreaks[1]
\newtheorem{theorem}{Theorem}
\newtheorem{lemma}{Lemma}

\newtheorem{corollary}{Corollary}

\newcommand{\Pe}{P_{\mathrm e}}
\newcommand{\JR}{M}
\newcommand{\Dm}{D}
\newcommand{\aR}{r_{\mathrm r}}
\newcommand{\Bin}{\operatorname{Bin}}
\newcommand{\Bern}{\operatorname{Bern}}

\newcommand{\KL}{D_{\mathrm{KL}}}
\newcommand{\E}{\mathbb E}

\newcommand{\Nstar}{N_\epsilon^\star}
\newcommand{\Ntot}{N_{\mathrm{tot}}}
\begin{document}
\title{A Particle-Based Multiuser Communication
with a Budget Limit}
\author{Shaojie Zhang,~\IEEEmembership{Student Member,~IEEE}
        and Ozgur B. Akan,~\IEEEmembership{Fellow,~IEEE}
        \thanks{The authors are with the Centre for neXt Communications (CXC),
        Department of Engineering, University of Cambridge, CB3 0FA Cambridge, UK.
        Their e-mail addresses are sz466@cam.ac.uk and oba21@cam.ac.uk.}
        \thanks{Ozgur B. Akan is also with the Centre for neXt Communications (CXC),
        Department of Electrical and Electronics Engineering, Ko\c{c} University,
        34450 Istanbul, T\"{u}rkiye. The e-mail address is akan@ku.edu.tr.}
}
\date{}
\maketitle
\begin{abstract}
Multiuser networks can aggregate local observations through a shared receiver. In this paper, we study the particle budget required to recover the exact sum of binary inputs in a multiuser network. We show that, for the considered model, the minimum budget grows quadratically with the number of transmitters when the symbol duration and target error probability are fixed, even when the receiver observes all particle arrival times.
Simultaneous on-off keying (OOK) achieves this quadratic scaling, while any slower budget growth makes the error probability approach one at a fixed symbol duration, regardless of the transmission and decoding scheme. We then examine how this particle requirement can be reduced by allowing longer symbol durations and more counting intervals. We develop a groupwise OOK counting (GOC) scheme with staggered group releases and sequential estimation of group input sums, accounting for delayed arrivals. Our bounds show that, when reception is based on interval counts, reducing the quadratic scaling requires both the symbol duration and the number of counting intervals to grow with network size. When both grow in proportion to network size, the minimum budget becomes proportional to the number of transmitters times its logarithm.

\end{abstract}


\section{Introduction}
\label{sec:intro}

Multiuser networks can combine local observations from several devices at a common receiver.
Depending on the task, the receiver may need only an aggregate of these observations rather than the individual values.
When particles carry the information, the devices communicate through particle releases and the receiver observes the resulting arrivals.
The number of particles released is then a natural measure of the communication cost.

Molecular communication provides a physical motivation for this setting, with molecules used as information carriers~\cite{farsad16survey,kuscu19}.
Diffusive channel models describe random arrival delays and possible nonarrival~\cite{yilmaz,jamali19}.
Such networks have been proposed for distributed sensing, including abnormality detection and localization~\cite{etemadi23}.

The receiver counts positive device observations without identifying the devices.
Received particles do not directly give this count: particles may arrive after the decision or never arrive~\cite{yilmaz,jamali19}, and a release for a positive observation may produce no absorption before the decision.
We ask how the particle budget for reliable counting grows with network size and whether longer durations or detailed arrival information can reduce it.

Cooperative abnormality detection produces a binary decision from sensor observations~\cite{mosayebi17}, while cooperative molecular detection combines information from several devices about a common transmitted symbol~\cite{fang16,fang17simple,fang19}. 
No exact number of positive observations is required by these detectors. 
In other schemes, molecules are utilized to represent values,
combine them, and thereby perform sums or arithmetic~\cite{long24,long26}. 
For binary observations, the receiver can compare an exact count with different decision thresholds or, when the number of devices is known, convert it to a fraction. 
Spatial multiuser molecular communication models have also been used to characterize the collective received signal when transmitters send a common bit sequence~\cite{deng17}. 

Strategy-independent point-to-point scaling laws relate molecule numbers and time~\cite{eckford}, while studies with molecule-release constraints address communication performance and resource design~\cite{ghavami16,chen21allocation,rudsari19tdma}. 

The interpretation of these resource limits depends on the physical input and output mechanisms, timing, memory, and the imposed constraints~\cite{gohari16,zhang26ion,zhang25receiver}. 
Communication theory addresses information carried by molecule release and arrival times~\cite{farsad,srinivas12,rose}, while capacity analysis addresses point-to-point and multiple-access ligand-receptor channels~\cite{aminian15}. 
When a sample is taken, all molecules that are within a passive receiver's observation volume are counted~\cite{noel14}. 
Point-to-point capacity and scaling results alone do not account for how many particles must be released when each transmitter knows only its own bit and the receiver must recover the exact integer sum from all absorption times by the deadline. 

We consider a particle-based multiuser network with one particle type and independent propagation delays drawn from a common distribution. 
The arrival distribution allows permanent nonarrival, and the results for all arrival times require a uniform comparison condition for nearby releases.
Diffusion from equidistant transmitters to a fully absorbing spherical receiver is one instance of this model.
The model assigns an independent equiprobable input bit to each transmitter, while each transmitter chooses its releases based only on that bit and without feedback, and the receiver estimates the exact integer sum. 
The receiver observes either absorbed-particle counts over predetermined intervals or all absorption times by the deadline, with neither observation identifying the transmitters. 
For a fixed error target, our lower bounds apply to every release rule allowed in Section~\ref{sec:model} and every corresponding decoder. 
The scaling results and strong converse are derived under either a maximum particle budget or a constraint only on the expected total number released. 
The main results are as follows. 
\begin{enumerate}
\item \textbf{Limits with a bounded number of counting intervals.}
If the number of predetermined counting intervals is bounded independently of network size, the minimum particle budget is, up to constant factors, quadratic in the number of transmitters and inversely proportional to the probability that a particle released at the start is absorbed by the deadline. 
The lower bound holds for every allowed release rule and decoder, whereas simultaneous on--off keying (OOK) attains the same order. 

\item \textbf{Exact counting from all arrival times.}
For every fixed positive deadline, the minimum particle budget is quadratic in the number of transmitters even when all absorption times are observed, and simultaneous OOK attains this order. 
For every allowed release rule and decoder, any budget growing more slowly than quadratically makes the error probability tend to one. 

\item \textbf{Joint resource tradeoffs.}
We determine the growth of the required particle budget for the case when the symbol duration and, for interval-count observations, the number of counting intervals remain fixed or grow polynomially with network size. 
The lower and upper bounds have the same polynomial growth with network size but retain a logarithmic gap in intermediate regimes. We develop a groupwise OOK counting (GOC) scheme and derive a nonasymptotic error bound that yields the upper bounds while accounting for delayed arrivals from earlier groups. 
\end{enumerate}

Section~\ref{sec:model} defines the model and costs; Sections~\ref{sec:finitebin} and~\ref{sec:full} establish the interval-count and full-record limits.
Sections~\ref{sec:ach} and~\ref{sec:numerical} present resource tradeoffs and numerical results, respectively.

\section{Particle Releases, Budgets, and Receiver Observations}
\label{sec:model}

The particle-based multiuser communication network comprises
$K$ transmitters (Txs), each with one input bit, and a common receiver (Rx). 
The receiver makes its decision about the number of input bits equal to $1$ based on the absorbed particles. 
All transmitters are modeled as point sources that release the same type of particles to carry information. 
All particles have independent propagation delays with a common distribution that does not depend on the transmitter or release time.
Propagation is independent of the input bits and release choices.
Each particle is recorded once upon its first arrival and removed from the medium, an event referred to as absorption.

 \subsection{Release rules and particle budgets}
\label{sec:reports-budgets}
\label{sec:local-codes}

The symbol interval has duration $T>0$. 
Transmitter $i$ has input bit $W_i\in\{0,1\}$. 
The bits are assumed to be independent and equiprobable. 
To send $w$, $N_{\mathrm{tx},i}(w)$ particles are released at $\tau_{i\ell}(w)\in[0,T]$, $\ell=1,\ldots,N_{\mathrm{tx},i}(w)$. 
Each transmitter chooses the number of particles and their release times deterministically or randomly, using a distribution dependent only on its own input bit. 
Given the input bits, these choices are assumed independent across transmitters and independent of particle transport. 

For each transmitter $i$, we require the particle budget $N_i$ to satisfy $N_{\mathrm{tx},i}(w)\le N_i$ for both inputs and every allowed release sequence. 
First, the network is subject to a particle-budget constraint $\Ntot$ such that
\begin{equation}
\sum_{i=1}^K N_i\le\Ntot .
\label{eq:budget}
\end{equation}
Alternatively, we constrain only the expected total number of particles released, with no maximum per symbol interval. 
We define the expected total number of particles released under release rules $\mathcal C$ as
\begin{equation}
B(\mathcal C)=
\E\left[\sum_{i=1}^K N_{\mathrm{tx},i}(W_i)\right].
\label{eq:expected-emission-cost}
\end{equation}
Averaging over the input bits and any random selection of release sequences allows this expectation to be noninteger. 
Releases at $t=T$ are counted under both criteria. 
The particle-budget constraint results in $B(\mathcal C)\le\Ntot$. 

\subsection{Particle arrivals and receiver observations}

Let us define $F(t)$ as the probability of a particle released by a transmitter at time zero being absorbed at the receiver by time $t$. 
Take $F(t)=0$ for $t\le0$, and assume $F(t)>0$ for every $t>0$.
The probability of eventual absorption is $p_\infty=\lim_{t\to\infty}F(t)\in(0,1)$, so each particle has a positive probability of never reaching the receiver.
The distribution $F$ is fixed as the network grows.

As a diffusion example, consider the following model.
The receiver is a fully absorbing sphere of radius $\aR$, and every transmitter lies at distance $r_0>\aR$ from the receiver's center. 
We assume that every particle diffuses independently with coefficient $\Dm$ in an unbounded, homogeneous three-dimensional medium. 
The medium initially contains no signalling particles and has only the receiver surface as a boundary. 
Flow, degradation, and interactions between signalling particles are considered negligible. 
Every particle that first contacts the receiver surface is absorbed and removed from the medium~\cite{yilmaz}. 
In the absorbing-sphere model, we have~\cite{yilmaz}
\begin{equation}
F(t)=\frac{\aR}{r_0}\operatorname{erfc}\!\left(
\frac{r_0-\aR}{\sqrt{4\Dm t}}\right),\quad t>0.
\label{eq:F}
\end{equation}
In this example, $p_\infty=\aR/r_0$. 
For the general arrival distribution, a particle released at $\tau\in[0,T]$ arrives at the receiver by the deadline with probability $F(T-\tau)$. 

All arrival times of particles absorbed by $T$ may serve as the receiver's observation. 
In $(0,T]$, this finite collection is denoted by $\mathcal Y_T$, including repeated arrival times. 
If there are no absorbed particles, then $\mathcal Y_T=\varnothing$. 
The particles are indistinguishable at the receiver, which records neither transmitter identities nor absorption locations. 

Let $N_{\mathrm{abs}}(t)$ be the total number of particles that are absorbed by time $t$. 
With predetermined boundaries $\boldsymbol t=(t_0,\ldots,t_{\JR})$ satisfying $0=t_0<\cdots<t_{\JR}=T$, the interval-count receiver observes
\begin{equation}
Y_j=N_{\mathrm{abs}}(t_j)-N_{\mathrm{abs}}(t_{j-1}),
\qquad j=1,\ldots,\JR.
\label{eq:physicalcount}
\end{equation}
The interval-count vector $\boldsymbol Y=(Y_1,\ldots,Y_{\JR})$ depends deterministically on $\mathcal Y_T$, and $\JR$ denotes the number of counting intervals. 

A particle released at $\tau\in[0,T]$ arrives at the receiver in counting interval $j$ with probability $F(t_j-\tau)-F(t_{j-1}-\tau)$, as in the diffusion model~\cite{yilmaz,deng17}. 
The probability of no absorption by $T$ is $1-F(T-\tau)$. 
The arrival outcomes are independent conditional on the released particles and their release times. 
The interval counts are generally dependent because each particle can be absorbed in at most one counting interval. 
We assume a common time reference and knowledge of the channel statistics at the transmitters and receiver~\cite{farsad,deng17}. 
Each transmitter's release rule and the distribution of any randomly selected release sequence are known to the receiver, but the selected sequence is not. 

A decoder estimates $U=\sum_{i=1}^K W_i$
as an integer $\widehat U$ between zero and $K$ based on either observation. 
We define the error probability as $\Pe=\Pr\{\widehat U\ne U\}$, averaging over the input bits, release sequences, and particle transport. 
The target probability of incorrectly estimating the total count is fixed at $\epsilon\in(0,1)$. 

\section{Particle-budget limits with interval counts}
\label{sec:finitebin}
\label{sec:count-budget}

Random particle arrivals can make input sums that differ by one difficult to distinguish.
We ask how the required particle budget grows with network size when the number of counting intervals remains bounded.

For predetermined counting intervals with boundaries $\boldsymbol t$, define $\Nstar(K,T;\boldsymbol t)$ as the minimum network particle budget over all release rules allowed by Section~\ref{sec:model} and all interval-count decoders achieving $\Pe\le\epsilon$. 

\subsection{Particle-budget lower bound with interval-count observations}
\label{sec:count-converse}

Consider a network of $K\ge2$ transmitters with particle budget $\Ntot$. 

\subsubsection{A conditional description of the received counts}
For each transmitter, sample the particle number and release times for input 0 and input 1 according to the corresponding release rules. 
Across transmitters and input values, the particle numbers and release times are independent, and they are independent of the input bits and particle propagation. 
Only the releases for the actual input bit occur, so the joint distribution of the input bits, actual releases, and received signal is unchanged. 
Because sampling is independent of the input bits, fixing both possibilities at every transmitter preserves the bits' independence and equiprobability. 

Each release-time sequence is represented by $N_i$ entries, where $\infty$
indicates that the corresponding particle is not released. 
The index $\ell$ denotes a sequence position rather than an observed particle identity. 
For each $(i,\ell)$, define $\Gamma_{i\ell}$ as the indicator of absorption by $T$ of a particle released at time zero, with probability $F(T)$, and $V_{i\ell}$ as its absorption delay conditional on absorption by $T$. 
We take these variables to be mutually independent and independent of the input bits and the particle numbers and release times specified for both input values. 
We can write the distributions of these variables as
\begin{equation}
\begin{gathered}
\Gamma_{i\ell}\sim\Bern(F(T)),\\
\Pr\{V_{i\ell}\le t\}=F(t)/F(T),\quad 0\le t\le T.
\end{gathered}
\end{equation} 
In this representation, a particle released at $\tau$ is absorbed and recorded at $\tau+V_{i\ell}$ if and only if $\Gamma_{i\ell}=1$ and $\tau+V_{i\ell}\le T$. 
For $\tau\le t\le T$, the corresponding absorption probability satisfies
$\Pr\{\Gamma_{i\ell}=1,\ \tau+V_{i\ell}\le t\}
=F(t-\tau)$. 

To obtain the lower bound, give the receiver the particle numbers and release times for both input values at each transmitter, together with $V_{i\ell}$, but not $\Gamma_{i\ell}$. 
With this information fixed, define $a_{ij}(w)$ as the number of particles transmitter $i$ would release for input $w$ at finite times $\tau_{i\ell}(w)$ satisfying $\tau_{i\ell}(w)+V_{i\ell}\in(t_{j-1},t_j]$. 
Conditioning on the fixed sequences and delays yields
$\sum_j a_{ij}(w)\le N_i$ and
\begin{equation}
Y_j\mid\boldsymbol W\sim
\Bin\!\left(\sum_i a_{ij}(W_i),F(T)\right).
\label{eq:condbin}
\end{equation} 
The interval counts depend on disjoint sets of $\Gamma_{i\ell}$ and are therefore independent given the input bits, the particle numbers and release times for both input values, and the auxiliary delays. 
Moreover, we denote these counts and the receiver's additional information by $\mathcal O$. 
Every error lower bound based on $\mathcal O$ consequently holds for the original receiver, which observes only the interval counts. 

\phantomsection\label{sec:adjacent-binomial}
\subsubsection{Comparing binomial count distributions}
Under~\eqref{eq:condbin}, changing a bit changes the binomial trial count at fixed success probability, whereas~\cite{chen} varies success probabilities at fixed trial count.
Use the unnormalized squared Hellinger distance $H^2$ and chi-squared divergence $\chi^2$~\cite[Sec.~II-B]{sasonverdu16}, and write $\mathsf B_n=\Bin(n,p)$ for $0<p<1$.
The mass ratio of $\mathsf B_n$ to $\mathsf B_{n+1}$ is $(n+1-y)/[(n+1)(1-p)]$; its variance under $\mathsf B_{n+1}$ gives
\begin{equation}
\chi^2(\mathsf B_n\Vert\mathsf B_{n+1})
=\frac{p}{(n+1)(1-p)}.
\label{eq:adjacent-binomial}
\end{equation}
For $1\le d\le m$ and $n\ge d$, the Hellinger triangle inequality along $\mathsf B_n,\ldots,\mathsf B_{n+d}$ gives
$H^2(\mathsf B_n,\mathsf B_{n+d})\le pd^2/[(1-p)(n+1)]$.
For $n<d$, contraction under convolution with $\mathsf B_n$ gives
$H^2(\mathsf B_n,\mathsf B_{n+d})\le2[1-(1-p)^{d/2}]\le-d\ln(1-p)\le pd/(1-p)$.
Using $d\le m$ and, respectively, $n+d\le2(n+1)$ or $n+d<2d$, both cases yield
\begin{equation}
H^2(\mathsf B_n,\mathsf B_{n+d})
\le \frac{2pmd}{(1-p)(n+d)}.
\label{eq:hell}
\end{equation}

\subsubsection{Summing the comparison bounds over transmitters}
Note that $\sum_i N_i\le\Ntot$ implies that at least half of the transmitters have particle budgets at most $2\Ntot/K$. 
Let $\mathcal I$ denote their index set. 

Fix the release sequences and the auxiliary delays. 
With $d_{ij}=|a_{ij}(1)-a_{ij}(0)|$ defined, write
\[
a_{ij}(W_i)=\min_w a_{ij}(w)+d_{ij}A_{ij},
\]
where $A_{ij}$ indicates whether the actual input gives the larger contribution to interval $j$. 
When $d_{ij}>0$, take $A_{ij}=W_i$ if $a_{ij}(1)>a_{ij}(0)$, and $A_{ij}=1-W_i$ otherwise. 
Independence and equiprobability of the input bits imply that the corresponding $A_{ij}$ are independent and equiprobable across transmitters within each fixed interval $j$. 
When $d_{ij}=0$, choose $A_{ij}=0$. 

We use
\[
S_j=\sum_{i\in\mathcal I}d_{ij}A_{ij}
\]
to denote the input-dependent part of the potential particle count from transmitters in $\mathcal I$. 

If $i\in\mathcal I$, all other transmitters' input bits are denoted by $\boldsymbol W_{-i}$. 
Holding these bits fixed, define $P_i^w$ as the conditional distribution of $\boldsymbol Y$ for $W_i=w$. 
To apply~\eqref{eq:hell}, remove the common binomial contributions from transmitters outside $\mathcal I$ and the $\min_w a_{kj}(w)$ terms. 
In interval $j$, the two remaining count distributions are $\mathsf B_{S_j-d_{ij}A_{ij}}$ and $\mathsf B_{S_j-d_{ij}A_{ij}+d_{ij}}$, in either order. 
The Hellinger distance cannot increase when the common independent counts are added back. 
Conditional independence of the interval counts makes the squared distance between the joint distributions no greater than the sum of the interval-wise squared distances. 
Applying~\eqref{eq:hell} with $m=N_i\le 2\Ntot/K$ gives
\[
H^2(P_i^0,P_i^1)
\le\frac{4F(T)\Ntot}{(1-F(T))K}
\sum_j\frac{d_{ij}}{S_j-d_{ij}A_{ij}+d_{ij}},
\]
where terms with $d_{ij}=0$ are defined to be zero. 

Fix an interval $j$ to sum the Hellinger bounds over transmitters. 
Since $S_j-d_{ij}A_{ij}$ is determined solely by contributions from other transmitters, the denominator is independent of $A_{ij}$. 
Averaging over the two equally likely values of $A_{ij}$ yields
\begin{equation}
\sum_{i\in\mathcal I}\E
\frac{d_{ij}}{S_j-d_{ij}A_{ij}+d_{ij}}
=2\Pr\{S_j>0\}\le2.
\label{eq:robust-sharing}
\end{equation} 
Applying this bound to the preceding Hellinger estimate, summing over the $\JR$ observation intervals, and averaging over both release sequences and the auxiliary delays gives
\begin{equation}
\sum_{i\in\mathcal I}\E H^2(P_i^0,P_i^1)
\le\frac{8F(T)\JR\Ntot}{(1-F(T))K}.
\label{eq:robust-shared-H}
\end{equation} 

\subsubsection{Recovering one bit}
Reveal all bits except that of a uniformly selected transmitter $i\in\mathcal I$.
Recovering $U$ then recovers $W_i$, whose binary testing error is at least $(1-H(P_i^0,P_i^1))/2$.
Averaging and using Cauchy--Schwarz, $|\mathcal I|\ge K/2$, and~\eqref{eq:robust-shared-H} gives
\begin{equation}
\Pe\ge\left[
\frac12-\frac{2}{K}\sqrt{\frac{F(T)\JR\Ntot}{1-F(T)}}
\right]_+,
\label{eq:r15-count-binary}
\end{equation}
where $[x]_+=\max\{x,0\}$.

\subsubsection{From received-count laws to an information bound}

To bound information about a hidden bit, write $\KL$ for relative entropy in nats and define $M_\theta=(1-\theta)P+\theta Q$ for probability laws $P,Q$ and $0\le\theta\le1$.
Then
\begin{equation}
 (1-\theta)\KL(P\Vert M_\theta)
 +\theta\KL(Q\Vert M_\theta)
 \le H^2(P,Q).
 \label{eq:robust-weighted-js}
\end{equation}
Indeed, $\KL\le\chi^2$ reduces the claim to
$\theta(1-\theta)(\sqrt r+\sqrt s)^2\le(1-\theta)r+\theta s$ for densities $r,s$; the right side minus the left is $((1-\theta)\sqrt r-\theta\sqrt s)^2\ge0$.

\phantomsection\label{sec:hidden-sum}
\subsubsection{A subset bound for exact counting}

Reveal all input bits except those of transmitters in a set $\mathcal A$. 
Exact recovery of $U$ therefore requires that the number of transmitters in $\mathcal A$ with input bit $1$ be determined. 
For any receiver observation $\mathcal O$, the counting error is bounded below by
\begin{equation}
\begin{aligned}
\Pe\ge{}&\Biggl[1-
\max_u\Pr\left\{\sum_{i\in\mathcal A}W_i=u\right\}\\
&\qquad-\sqrt{\frac12\sum_{i\in\mathcal A}
I(W_i;\mathcal O\mid\boldsymbol W_{-i})}
\Biggr]_+ .
\end{aligned}
\label{eq:hidden-sum}
\end{equation} 
Without information about these transmitters, the receiver can do no better than guess their most likely number of positive reports. 
Pinsker's inequality bounds the improvement due to $\mathcal O$, while the chain rule and independence of the input bits yield the sum in \eqref{eq:hidden-sum}, as in the conditional-information argument of~\cite{xu}. 

For the enhanced interval-count observation above, select in advance the $\ell$ transmitters in $\mathcal I$ for which $\mathcal O$ gives the least conditional information about their bits, and denote their set by $\mathcal A$, where $1\le\ell\le\lfloor K/2\rfloor$. 
Using $|\mathcal I|\ge K/2$, we combine \eqref{eq:robust-weighted-js} and \eqref{eq:robust-shared-H} to obtain
\[
\sum_{i\in\mathcal A}
I(W_i;\mathcal O\mid\boldsymbol W_{-i})
\le
\frac{16F(T)\ell\JR\Ntot}
{(1-F(T))K^2}.
\] 
For these independent equiprobable inputs, let
\[
\beta_\ell
=
2^{-\ell}\binom{\ell}{\lfloor\ell/2\rfloor},
\]
be the probability of the most likely number of positive reports, and let $\beta_0=1$. 
Substituting these bounds into \eqref{eq:hidden-sum} and removing the auxiliary release sequences and delays available only to the enhanced receiver yields
\begin{equation}
\Pe\ge\left[
1-\beta_\ell-
\sqrt{\frac{8F(T)\ell\JR\Ntot}
{K^2(1-F(T))}}
\right]_+,
\quad
1\le\ell\le\lfloor K/2\rfloor.
\label{eq:subset}
\end{equation} 
For every predetermined partition into a finite
number $\JR$ of counting intervals, this bound holds. 

\subsection{Particle-budget scaling and expected number of particles released}
\label{sec:count-scaling}

We combine the preceding lower bound with a simultaneous OOK construction to bound the minimum particle budget.

\begin{theorem}[Particle-budget scaling with a bounded number of sampling intervals]
\label{thm:budget-duration}
Fix the error target $\epsilon\in(0,1)$ together with an integer $M_0\ge1$. 
Constants $c,C>0$ and an integer $K_0$ exist, depending only on $\epsilon,M_0,p_\infty$, for which
\begin{equation}
c\,\frac{K^2}{F(T)}
\le\Nstar(K,T;\boldsymbol t)
\le C\,\frac{K^2}{F(T)}
\label{eq:budget-order}
\end{equation}
holds for every $K\ge K_0$, $T>0$, and every set of predefined observation intervals in $[0,T]$ containing between one and $M_0$ sampling intervals. 
The constants and $K_0$ do not depend on the symbol duration or interval boundaries. 
\end{theorem}

\begin{IEEEproof}
Note that the lower bound in Theorem~\ref{thm:budget-duration} follows from \eqref{eq:subset}
because we choose a fixed $\ell$ with $\beta_\ell\le(1-\epsilon)/2$
and use $\JR\le M_0$ and $1-F(T)\ge1-p_\infty>0$. 

For the upper bound, transmitter $i$ uses simultaneous local OOK to release $NW_i$ particles at $t=0$. 
The total absorbed count, $N_{\mathrm{abs}}(T)=\sum_{j=1}^{\JR}Y_j$, satisfies $N_{\mathrm{abs}}(T)\mid U=u\sim\Bin(Nu,F(T))$~\cite{deng17}. 
The receiver estimates $U$ by selecting the integer in $\{0,\ldots,K\}$ closest to $N_{\mathrm{abs}}(T)/(NF(T))$. 
Using Chebyshev's argument in \cite[App.~B]{eckford}, we bound the binomial counting error under $u\le K$ and obtain
\begin{equation}
\Pe\le\frac{4K(1-F(T))}{NF(T)}.
\end{equation} 
By setting $N=\lceil4K(1-F(T))/(\epsilon F(T))\rceil$, we ensure that $\Pe\le\epsilon$ and obtain
\[
N_\epsilon^\star(K,T;\boldsymbol t)
\le KN
\le(4/\epsilon+1)K^2/F(T).
\] 
Since the receiver uses only the total absorbed count by time $T$, the bound holds regardless of counting-interval number and placement. 

\end{IEEEproof}

Let $B_\epsilon^\star(K,T;\boldsymbol t)$ be the infimum of~\eqref{eq:expected-emission-cost} over the allowed release rules and interval-count decoders with $\Pe\le\epsilon$, without a per-symbol cap.
Theorem~\ref{thm:budget-duration} also holds with $B_\epsilon^\star$ replacing $\Nstar$.
To see this, fix both input-dependent release sequences at every transmitter and set
$N_{\mathrm{cap}}=\sum_i\max_w N_{\mathrm{tx},i}(w)$.
The fair prior gives $\E N_{\mathrm{cap}}\le2B(\mathcal C)$.
Applying~\eqref{eq:subset} conditionally with this cap, then averaging and using Jensen, yields
\begin{equation}
\begin{gathered}
1-\Pe\le\beta_\ell+
\frac{\sqrt{16F(T)\ell\JR B(\mathcal C)/(1-F(T))}}{K},\\
1\le\ell\le\lfloor K/2\rfloor.
\end{gathered}
\label{eq:expected-count-success}
\end{equation}
The same fixed $\ell$ gives the lower bound; simultaneous OOK gives the upper bound with average cost $KN/2$~\cite{gursoy19}.

\section{Duration limits with all arrival times}
\label{sec:full}

The preceding lower bounds apply to interval counts.
Arrival times within each interval may contain additional information when transmitters use different release times.
We ask whether observing all arrival times can change the growth order of the required particle budget with network size at a fixed symbol duration. 

Let $N_{\epsilon,\mathrm{full}}^\star(K,T)$ denote the minimum
particle budget required under the model of Section~\ref{sec:model}
for a decoder based on all arrival times to achieve $\Pe\le\epsilon$. 

\subsection{Comparing distributions of arrival times}
\label{sec:record-insertion}

The arrival records up to $T$ are compared with and without
one additional particle released at $\tau\in[0,T]$. 
We denote this particle's record distribution by $R_{\tau,T}$. 

For the results in this section, assume that fixed constants $\tau_0>0$ and $\alpha_0\in(0,1]$ satisfy
\[
R_{\sigma,T}\ge\alpha_0R_{\tau,T}
\quad\text{if}\quad
0\le\sigma\le\tau\le T,\quad \tau-\sigma\le\tau_0,
\]
for every $T>0$.
The inequality applies to every measurable set of records, including no arrival by $T$.
Thus, arrival-record distributions for releases within a window of width $\tau_0$ share a common component of weight $\alpha_0$.

A sufficient condition is that the finite arrival distribution has a density $f$ with $F(t)=\int_0^t f(s)\,ds$ and
\[
f(t+h)\ge\alpha_0 f(t),\qquad 0\le h\le\tau_0,
\]
for each $h$ and almost every $t>0$, where $0<\alpha_0\le1-p_\infty$.
The density inequality compares the observed arrivals, while the probability of no arrival for the earlier release is at least $1-p_\infty$ and that for the later release is at most one.

For diffusion, set $\tau_{\mathrm d}=(r_0-\aR)^2/(4\Dm)$.
Differentiating~\eqref{eq:F}~\cite{yilmaz,tepekule} gives
\[
(\log F')'(t)=\frac{\tau_{\mathrm d}}{t^2}-\frac{3}{2t}
\ge-\frac{9}{16\tau_{\mathrm d}},\qquad t>0.
\]
Integration over length at most $\tau_{\mathrm d}$ gives the density comparison; including no arrival, take $\tau_0=\tau_{\mathrm d}$ and $\alpha_0=\min\{1-p_\infty,e^{-9/16}\}$.
The same condition also holds for $F(t)=p_\infty(1-e^{-\lambda t})$ with $\lambda>0$, by taking $\tau_0=1/\lambda$ and $\alpha_0=\min\{1-p_\infty,e^{-1}\}$.

For the general arrival distribution, define
\begin{equation}
C_{\mathrm{win}}(T)
=\frac{F(T)}{\alpha_0[1-F(T)]}.
\label{eq:window-constants}
\end{equation} 
The factor $C_{\mathrm{win}}(T)$ captures both this overlap and
the probability of absorption by the deadline and controls how
an additional particle affects the received record. 

\begin{lemma}[Single-particle insertion within a release window]
\label{lem:same-window-insertion}
Under the preceding comparison condition, consider a symbol interval $[0,T]$ with $n$ independent particles released at times $\sigma_r$ satisfying $0\le\sigma_r\le\tau\le T$ and $\tau-\sigma_r\le\tau_0$, $r=1,\ldots,n$. 
Let $\mathsf P$ and $\mathsf Q$ denote, respectively, the distributions
of the combined arrival records up to $T$, without and
with one additional independent particle released at $\tau$. 
Both records can include any common independent component. 
This comparison leads to
\begin{equation}
 H^2(\mathsf P,\mathsf Q)
 \le\frac{C_{\mathrm{win}}(T)}{n+1}.
 \label{eq:same-window-insertion}
\end{equation} 
\end{lemma}

\begin{IEEEproof}
Write $p=F(T-\tau)$. 
If $p=0$, then $\mathsf P=\mathsf Q$. Otherwise, assume $p>0$. 

By $R_{\sigma_r,T}\ge\alpha_0R_{\tau,T}$, each existing
record distribution can be written as a mixture of $R_{\tau,T}$ with
weight $\alpha_0$ and a residual distribution. 

For this comparison only, the receiver is given the independent mixture
selections and residual records, which have the same joint distribution
in both cases~\cite[Sec.~III, Theorem~III.2
and Lemma~III.3]{fmt}. 

Let $L\sim\Bin(n,\alpha_0)$ be the number of records from $R_{\tau,T}$ that are selected among the $n$ particles. 
Once the residual records are set aside, we combine the remaining records without particle or transmitter labels. 
The absorption counts conditioned on the revealed information are $\Bin(L,p)$ and $\Bin(L+1,p)$, respectively. 

Given the same count, the conditional distribution of the arrival times
is identical in both cases. 

After applying~\eqref{eq:adjacent-binomial} and $H^2\le\chi^2$
to the absorption counts, we average over the revealed information
and add the remaining independent arrival records common to both cases.
Neither operation increases the Hellinger distance, so we obtain 
\begin{equation}
\begin{aligned}
H^2(\mathsf P,\mathsf Q)
&\le \frac{p}{1-p}\,
     \E\left[\frac{1}{L+1}\right]\\
&\le \frac{p}{\alpha_0(1-p)(n+1)}
 \le \frac{C_{\mathrm{win}}(T)}{n+1},
\end{aligned}
\label{eq:insertion-smoothing}
\end{equation}
where we used the inequalities $\E[(L+1)^{-1}]\le[\alpha_0(n+1)]^{-1}$ and $p\le F(T)$. 
\end{IEEEproof}

\subsection{Release-order analysis and duration bounds}
\label{sec:record-aggregation}

\begin{theorem}[Particle-budget and duration bounds]
\label{thm:full-duration}
Under the comparison condition of Subsection~\ref{sec:record-insertion}, fix \(\epsilon\in(0,1)\).
There exist a constant \(c_\epsilon>0\), depending only on \(\epsilon,\alpha_0,p_\infty\), and an integer
\(K_\epsilon\), neither of which depends on \(T\) or the network
particle budget \(\Ntot\). 

For all release rules permitted by Section~\ref{sec:model}
and every decoder based on all arrival times,
\(K\ge K_\epsilon\) and \(\Pe\le\epsilon\) imply
\begin{equation}
\Ntot F(T)
\left(1+\frac{T}{\tau_0}\right)
\ge
c_\epsilon K^2.
\label{eq:full-duration}
\end{equation} 

The release rules and decoder are allowed to depend on network size. If
\begin{equation}
N_{\mathrm{tot},K}
=o\left(
\frac{K^2}
{F(T_K)(1+T_K/\tau_0)}
\right),
\label{eq:full-duration-strong}
\end{equation}
then the probability of incorrect counting tends to one,
$\Pe\to1$. 
\end{theorem}

\begin{IEEEproof}[Proof of Theorem~\ref{thm:full-duration}]
We partition particle releases over $[0,T]$ into at most
$1+T/\tau_0$ consecutive windows, each of width
$\tau_0$ except for a shorter final window if needed. 
Let $\mathcal O_{\mathrm{win}}$ denote the collection of all arrival times
up to $T$, together with both release sequences at every
transmitter drawn as in Subsection~\ref{sec:count-converse},
where arrival times are labelled only by release window. 
The record retains arrivals after their release window closes. 

With these sequences and $\boldsymbol W_{-i}$ fixed, $P_i^w$ is
the distribution of $\mathcal O_{\mathrm{win}}$ given $W_i=w$. 
For each input value, we start with the other transmitters' arrivals
and add, one at a time in release-time order, each absorption time,
if any, from transmitter $i$'s particles. 
Let $n_{i\ell}(w)$ be the number of particles from other
transmitters released
in the same window by time $\tau_{i\ell}(w)$, together with
transmitter $i$'s particles already added there. 
The remaining particles contribute independently, and their arrival-record
distributions are unchanged. 
Using Lemma~\ref{lem:same-window-insertion} and the Hellinger
triangle inequality, we obtain
\begin{equation}
H(P_i^0,P_i^1)
\le\sqrt{C_{\mathrm{win}}(T)}
\sum_{w=0}^{1}\sum_{\ell=1}^{N_{\mathrm{tx},i}(w)}
[n_{i\ell}(w)+1]^{-1/2}.
\label{eq:window-codeword-H}
\end{equation} 

For $W_i$ equally likely to be $0$ or $1$, the expected
contribution from the selected sequence is one half of the sum
in~\eqref{eq:window-codeword-H} over both release sequences. 
To bound the contribution from each release window,
we order its particles by release time and keep the same
transmitter's insertion order when release times coincide. 

Consider the particle ranked $k$ in this ordering, released by transmitter $i$. 
Of the preceding $k-1$ particles, those from other
transmitters were released no later than this particle, whereas
those from transmitter $i$ had already been added. 
Both groups are included in $n_{i\ell}(W_i)$. Hence
$n_{i\ell}(W_i)+1\ge k$. 

$N_{\mathrm{win},r}$ is the number of particles
released in window $r$. 
Thus, that window contributes at most
$\sum_{k=1}^{N_{\mathrm{win},r}}k^{-1/2}
\le2\sqrt{N_{\mathrm{win},r}}$. 
By summing over all transmitters and averaging over the
input bits and release sequences, we obtain
\begin{equation}
\begin{aligned}
\sum_i\E H(P_i^0,P_i^1)
&\le4\sqrt{C_{\mathrm{win}}(T)}
     \E\sum_r\sqrt{N_{\mathrm{win},r}}\\
&\le4\sqrt{
C_{\mathrm{win}}(T)
\left(1+\frac{T}{\tau_0}\right)
\E\sum_rN_{\mathrm{win},r}}.
\end{aligned}
\label{eq:window-total-H}
\end{equation} 
The expectation on the left averages the other bits and
both release sequences, whereas those on the right include
all bits. 
Using Cauchy--Schwarz, the bound on the number of windows,
and Jensen, we obtain the second inequality. 

Using~\eqref{eq:robust-weighted-js} and $H^2\le\sqrt2H$,
we obtain
\[
\sum_i I(W_i;\mathcal O_{\mathrm{win}}\mid\boldsymbol W_{-i})
\le4\sqrt{
2C_{\mathrm{win}}(T)
\left(1+\frac{T}{\tau_0}\right)\Ntot}.
\] 
Once all averages are taken, we fix, for $1\le\ell\le K$,
the $\ell$ transmitters with the smallest conditional mutual
information values. 
The sum of these values is at most $\ell/K$ times this bound. 
Using~\eqref{eq:hidden-sum} and this bound, we obtain the
following bound on counting success, also for the original receiver,
\begin{equation}
1-\Pe
\le
\beta_\ell+
\sqrt{
\frac{2\ell}{K}
\sqrt{
2C_{\mathrm{win}}(T)
\left(1+\frac{T}{\tau_0}\right)\Ntot}
}.
\label{eq:release-window-success-fair}
\end{equation} 

Fix $\ell_\epsilon$ satisfying
$\beta_{\ell_\epsilon}\le(1-\epsilon)/2$. 
If $K\ge\ell_\epsilon$ and $\Pe\le\epsilon$, then
\eqref{eq:release-window-success-fair}, the definition of
$C_{\mathrm{win}}(T)$, and $1-F(T)\ge1-p_\infty$ give
\eqref{eq:full-duration} with
\[
c_\epsilon
=\frac{\alpha_0(1-p_\infty)(1-\epsilon)^4}
{128\ell_\epsilon^2}.
\] 

Using~\eqref{eq:full-duration-strong} and the same bound
on $1-F(T_K)$, we obtain
\[
C_{\mathrm{win}}(T_K)
\left(1+\frac{T_K}{\tau_0}\right)
N_{\mathrm{tot},K}=o(K^2).
\] 
For fixed $\ell$,~\eqref{eq:release-window-success-fair}
yields $\limsup_{K\to\infty}(1-\Pe)\le\beta_\ell$. 
As $\ell\to\infty$, we obtain $\Pe\to1$. 
\end{IEEEproof}

By the data-processing inequality, the conclusions also hold after the observed arrival times are postprocessed, provided that the postprocessing kernel uses only those observations and does not alter particle release, propagation, or absorption. 

Let $B_{\epsilon,\mathrm{full}}^\star(K,T)$ be the infimum of $B(\mathcal C)$ over all allowed release rules and full-record decoders with $\Pe\le\epsilon$, without a per-symbol cap.

\begin{corollary}[Expected cost and fixed-deadline scaling]
\label{cor:full-quadratic}
\label{cor:expected-emissions}
Theorem~\ref{thm:full-duration}, including its duration-dependent bound and strong converse, remains valid with $B(\mathcal C)$ replacing $\Ntot$.
For fixed $T>0$ and $\epsilon\in(0,1)$ under the same arrival conditions,
\begin{equation}
N_{\epsilon,\mathrm{full}}^\star(K,T)
=\Theta(K^2),\qquad
B_{\epsilon,\mathrm{full}}^\star(K,T)=\Theta(K^2).
\label{eq:full-quadratic}
\end{equation}
Under either cost constraint, subquadratic cost at this deadline implies $\Pe\to1$.
\end{corollary}
\begin{IEEEproof}
Each released particle belongs to exactly one release window, so
$\E\sum_rN_{\mathrm{win},r}=B(\mathcal C)$.
Substituting in~\eqref{eq:window-total-H} and following the same information and hidden-subset argument proves the expected-cost bound and strong converse.
At fixed $T$, Theorem~\ref{thm:full-duration} and this extension give both lower bounds; simultaneous OOK in Theorem~\ref{thm:budget-duration} gives both matching upper bounds.
\end{IEEEproof}

\section{Tradeoffs among particle budget, duration, and counting intervals}
\label{sec:ach}

Longer symbol intervals and more counting intervals can reduce the
required particle budget, but permanent nonabsorption imposes an
additional lower bound.

\subsection{Particle-budget lower bounds from nonarrival}
\label{sec:invisibility}
The variables $\Gamma_{i\ell}\sim\operatorname{Bern}(F(T))$ are independent, with $\Gamma_{i\ell}=1$ indicating that the
corresponding particle would be absorbed by the deadline if released at time zero. 
We can write the probability of all $N_i$ variables assigned to transmitter $i$ being zero as
\[
p_{\mathrm{inv},i}=(1-F(T))^{N_i}.
\] 
When $N_i=0$, this probability is one. 
Nonnegative release times ensure that, on this event, neither input value can
produce absorption by the deadline, regardless of the release times. 

We use $K_{\mathrm{inv}}$ to denote the number of transmitters that have no nonzero assigned variables. 
We give a decoder these transmitters' indices, all other bits, and the physical observation. 
Since these transmitters produce no arrivals and their indices are independent
of the inputs, the bits that remain unrevealed follow their independent prior distributions. 
For equiprobable bits, their sum is conditionally distributed as $\Bin(K_{\mathrm{inv}},0.5)$, giving
\begin{equation}
\Pe\ge1-\E[\beta_{K_{\mathrm{inv}}}].
\label{eq:invis}
\end{equation} 

Since the nonarrival events are independent, we have
\begin{equation}
\begin{aligned}
\Pe&\ge\frac12\Pr\{K_{\mathrm{inv}}\ge1\}\\
&=\frac12\left[1-\prod_{i=1}^{K}(1-p_{\mathrm{inv},i})\right]\\
&\ge\frac12\left[1-\{1-(1-F(T))^{\Ntot/K}\}^K\right].
\end{aligned}
\label{eq:simple-nonarrival}
\end{equation} 
The last inequality follows by applying Jensen's inequality to the concave, increasing
function $b\mapsto\ln(1-(1-F(T))^b)$ for $b>0$, since $\sum_i N_i\le\Ntot$. 

For $\Pe\le\epsilon<1/2$, inverting~\eqref{eq:r15-count-binary} and~\eqref{eq:simple-nonarrival} gives
\begin{equation}
\begin{gathered}
\Ntot \ge\max\Bigg\{\begin{gathered}
\frac{(1-2\epsilon)^2(1-F(T))K^2}{16F(T)\JR},\\
\frac{K}{-\ln(1-F(T))}
\ln\frac1{1-(1-2\epsilon)^{1/K}}
\end{gathered}\Bigg\}.
\label{eq:necessary}
\end{gathered}
\end{equation}

For independent $W_i\sim\Bern(q_i)$ with $q_i\in[\rho,1-\rho]$, $\rho\in(0,1/2]$, write $B_{\boldsymbol q}(\mathcal C)$ for the expected release cost under this prior.

\begin{lemma}[Nonarrival and the particle-cost floor]
\label{lem:uniform-invisibility-floor}
Set $\lambda_T=-\ln(1-F(T))$ and $\mu_0=K e^{-\lambda_T\Ntot/K}$.
If $\mu_0\ge2$, then
\begin{equation}
1-\Pe\le\frac4{\mu_0}
+\sqrt{\frac{\pi}{4\rho(1-\rho)\mu_0}}.
\label{eq:invisibility-envelope}
\end{equation}
For fair inputs, $\Pe\to1$ if $\lambda_{T_K}N_{\mathrm{tot},K}/K-\ln K\to-\infty$.
For fixed $\eta\in(0,1)$, $\rho$, and $p_\infty$, success probability at least $\eta$ requires, for all sufficiently large $K$,
\begin{align}
\Ntot&\ge cK\ln K,\label{eq:invisibility-hard-floor}\\
B_{\boldsymbol q}(\mathcal C)&\ge cK\ln K,\label{eq:invisibility-expected-floor}
\end{align}
under the respective constraints, with $c>0$ depending only on $(\eta,\rho,p_\infty)$.
These bounds hold uniformly over all allowed release rules, both observation models and their decoders, every $T>0$, and the stated priors.
\end{lemma}
\begin{IEEEproof}
Conditional on the invisible indices, $k\ge1$ unrevealed bits have total variance at least $k\rho(1-\rho)$; their sum has largest point mass at most $\sqrt{\pi/[8k\rho(1-\rho)]}$~\cite[Thm.~2.1]{bcv}.
Independence and Jensen give
$\mu=\E K_{\mathrm{inv}}=\sum_i e^{-\lambda_TN_i}\ge\mu_0$ and
$\operatorname{Var}(K_{\mathrm{inv}})\le\mu$.
Splitting at $K_{\mathrm{inv}}=\mu/2$ and using Chebyshev gives
$1-\Pe\le4/\mu+\sqrt{\pi/[4\rho(1-\rho)\mu]}$, proving~\eqref{eq:invisibility-envelope} and its strong-converse consequence.

Let $L=K\ln K/[-2\ln(1-p_\infty)]$.
Since $\lambda_T\le-\ln(1-p_\infty)$, $\Ntot\le L$ implies $\mu_0\ge\sqrt K$ and hence uniform success at most
$a_K=4K^{-1/2}+\sqrt{\pi/[4\rho(1-\rho)]}K^{-1/4}\to0$.
Thus fixed success requires $\Ntot>L$ for large $K$.

For expected cost, fix both release sequences as before.
The prior bounds give $\E N_{\mathrm{cap}}\le B_{\boldsymbol q}(\mathcal C)/\rho$.
The preceding uniform bound applies whenever $N_{\mathrm{cap}}\le L$; otherwise success is at most one.
Averaging and using Markov,
$1-\Pe\le a_K+B_{\boldsymbol q}(\mathcal C)/(\rho L)$.
For large $K$ with $a_K\le\eta/2$, success at least $\eta$ therefore requires $B_{\boldsymbol q}(\mathcal C)\ge\rho\eta L/2$.
\end{IEEEproof}

\subsection{Groupwise OOK counting with delayed arrivals}
\label{sec:grouped-ook}
The GOC scheme combines simultaneous transmissions within each group with staggered transmissions across groups.
The receiver sequentially estimates the group input sums using decision feedback to account for delayed arrivals, then adds these estimates to recover the total count.

Choose $\Delta>0$ as the interval width and define $t_j=j\Delta$,
$j=0,\ldots,\JR$, so that $T=\JR\Delta$. 
Divide the transmitters into $G=\min\{\JR,K\}$ nonempty groups,
with at most $K_{\mathrm{grp}}=\lceil K/G\rceil$ transmitters per group. 
We use $U_g$ to denote the number of input bits in group $g$ that equal 1. 
At time $(g-1)\Delta$, each transmitter in group $g$ releases
$N$ particles for bit 1 and remains silent for bit 0. 

For $\ell=0,1,\ldots$, a particle's probability of absorption between
$\ell\Delta$ and $(\ell+1)\Delta$ after release is~\cite{tepekule}
\begin{equation}
h_\ell=F((\ell+1)\Delta)-F(\ell\Delta),
\qquad h_0=F(\Delta)>0.
\end{equation} 
Given $\boldsymbol W$, each $Y_j$, $1\le j\le G$, sums independent particle indicators with absorption probabilities $h_{j-g}$~\cite{tepekule}.
Consequently,
\begin{equation}
\begin{aligned}
\E[Y_j\mid\boldsymbol W]&=N\sum_{g=1}^j h_{j-g}U_g,\\
\operatorname{Var}(Y_j\mid\boldsymbol W)&\le\E[Y_j\mid\boldsymbol W]
\le NK_{\mathrm{grp}}p_\infty,
\end{aligned}
\end{equation}
since $U_g\le K_{\mathrm{grp}}$ and $\sum_{\ell=0}^{j-1}h_\ell=F(j\Delta)\le p_\infty$.

For $j=1,\ldots,G$, the receiver subtracts the mean delayed-arrival contribution
from all earlier groups using their estimates~\cite{tepekule,noel}, before applying
\begin{equation}
\widehat U_j=
\operatorname*{arg\,min}_{u}
\left|Y_j-N\sum_{g=1}^{j-1}h_{j-g}\widehat U_g-Nh_0u\right|.
\end{equation} 
Only integers from zero to the number of transmitters in group $j$
need be considered as candidates. 
The receiver ignores any remaining intervals and outputs $\widehat U=\sum_{j=1}^{G}\widehat U_j$. 

For fixed inputs, if all first $G$ counts deviate from their conditional means by less than $Nh_0/2$, induction gives every group count correctly: correct earlier estimates leave the current residual nearest to $Nh_0U_j$.
Apply Bernstein's inequality~\cite[Cor.~2.11]{blm} at threshold $Nh_0/2$ to each count using the variance bound, take a union bound over the $G$ intervals, and average over inputs.
This requires neither independent interval counts nor conditioning on earlier correct decisions and yields
\begin{equation}
\Pe\le2G\exp\!\left[
-\frac{Nh_0^2}{8K_{\mathrm{grp}}p_\infty+4h_0/3}
\right].
\end{equation} 
The condition $\Pe\le\epsilon$ is ensured by choosing
\begin{equation}
N=\left\lceil
\frac{8K_{\mathrm{grp}}p_\infty+4h_0/3}{h_0^2}
\ln\frac{2G}{\epsilon}
\right\rceil.
\label{eq:grouped-ook-dose}
\end{equation} 

The network particle budget equals $KN$, whereas the expected number
of particles released equals $KN/2$ for equiprobable inputs. 
For fixed $\Delta$ and $\epsilon$,
\eqref{eq:grouped-ook-dose} yields
$KN=O(K\lceil K/G\rceil[1+\ln G])$. 
With $T$ fixed and $G$ growing, $h_0=F(T/\JR)$ varies,
so the fixed-$\Delta$ constants no longer apply. 

\subsection{Bounds on resource scaling}
\label{sec:resource-exponents}

$\Nstar(K,T,\JR)$ for interval-count observations is defined as the
minimum of $\Nstar(K,T;\boldsymbol t)$ over all choices of
$\JR$ positive-width counting intervals covering $[0,T]$. 
$B_\epsilon^\star(K,T,\JR)$ is defined similarly for the expected
number of particles released. 
Since the network particle budget bounds the expected number
of particles released, we have
$B_\epsilon^\star(K,T,\JR)\le\Nstar(K,T,\JR)$. 

Shortening GOC's counting intervals reduces
$F(\Delta)$, the probability that a particle is absorbed within
$\Delta$ after release, and may increase the particles needed
to achieve $\Pe\le\epsilon$. 
We therefore set a fixed counting-interval width
$\Delta_0=T_{\min}/2$. 
The absorption probability $F(\Delta_0)>0$ does not
depend on duration for all $T\ge T_{\min}$. 
For interval-count observations, we choose the number of
transmitter groups as
\begin{equation}
G(K,T,\JR)=
\begin{cases}
1,&\JR=1,\\[1mm]
\min\!\left\{
K,\JR-1,
\left\lceil T/\Delta_0\right\rceil-1
\right\},&\JR\ge2.
\end{cases}
\label{eq:bridge-G}
\end{equation}
Then, we choose the number of groups for observations of all arrival times as
\begin{equation}
G_{\mathrm{full}}(K,T)
=
\min\!\left\{
K,
\left\lfloor T/\Delta_0\right\rfloor
\right\}.
\label{eq:bridge-G-full}
\end{equation}
 \begin{theorem}[Three-resource bounds]
\label{thm:three-resource}
Take fixed $\epsilon\in(0,1)$, $T_{\min}>0$, and an arrival distribution satisfying Section~\ref{sec:model} and the comparison condition of Subsection~\ref{sec:record-insertion}, with fixed $\tau_0,\alpha_0$. 
Given these fixed quantities, there exist $c,C>0$ and
an integer $K_0\ge2$, depending only on them,
such that the following bounds hold for every $K\ge K_0$,
$T\ge T_{\min}$ and positive integer $\JR$. 
\begin{align}
B_\epsilon^\star(K,T,\JR)
&\ge c\max\Bigg\{
\begin{gathered}
\frac{K^2}{\JR F(T)},\ \frac{K^2}
{F(T)(1+T/\tau_0)},\\
K\ln K
\end{gathered}\Bigg\},
\label{eq:bridge-count-lower}\\
\Nstar(K,T,\JR)
&\le{}
CK\left\lceil\frac{K}{G(K,T,\JR)}\right\rceil\nonumber\\
&\quad\times[1+\ln G(K,T,\JR)].
\label{eq:bridge-count-upper}
\end{align}
When all arrival times are observed,
\begin{align}
B_{\epsilon,\mathrm{full}}^\star(K,T)
&\ge c\max\left\{
\frac{K^2}
{F(T)(1+T/\tau_0)},\
K\ln K
\right\},
\label{eq:bridge-full-lower}\\
N_{\epsilon,\mathrm{full}}^\star(K,T)
&\le{}
CK\left\lceil
\frac{K}{G_{\mathrm{full}}(K,T)}
\right\rceil
[1+\ln G_{\mathrm{full}}(K,T)].
\label{eq:bridge-full-upper}
\end{align}
\end{theorem}

\begin{IEEEproof}
The three terms in~\eqref{eq:bridge-count-lower} are obtained,
respectively, from~\eqref{eq:expected-count-success} with a fixed subset
size, Corollary~\ref{cor:expected-emissions}, and
Lemma~\ref{lem:uniform-invisibility-floor}. 
Since the last two bounds apply to all arrival times,
we can conclude that they apply to interval counts, too. 
Taking infima yields both lower bounds because these bounds
hold uniformly over the allowed release rules, decoders,
interval boundaries, and deadlines. 

We choose $G=G(K,T,\JR)$ for interval counts and $G=G_{\mathrm{full}}(K,T)$ for full records.
For $G=1$, simultaneous OOK uses only the total absorbed count and costs at most $(4/\epsilon+1)K^2/F(T_{\min})$.
For $G\ge2$, use GOC on the first $G$ intervals of width $\Delta_0$.
In the count model, $G\le\JR-1$ and $G\le\lceil T/\Delta_0\rceil-1$ ensure $G\Delta_0<T$; divide $(G\Delta_0,T]$ into $\JR-G$ positive-width intervals.
For full records, $G\Delta_0\le T$ suffices: form the first $G$ counts from arrival times and ignore later arrivals.
With at most $\lceil K/G\rceil$ transmitters per group and fixed $h_0=F(\Delta_0)>0$,~\eqref{eq:grouped-ook-dose} ensures $\Pe\le\epsilon$ at cost $O(K\lceil K/G\rceil[1+\ln G])$.
Choosing $C$ to cover both cases proves the upper bounds.
\end{IEEEproof}

Since $B_\epsilon^\star\le N_\epsilon^\star$ for either receiver observation,
the lower bounds apply to particle budgets and the upper bounds
to expected numbers of particles released. 

For fixed constants $c_t,c_m>0$ and exponents $b,m\ge0$, we set
\begin{equation}
T_K=c_t\tau_0K^b,
\qquad
\JR_K=\left\lceil c_mK^m\right\rceil.
\label{eq:polynomial-resource-scaling}
\end{equation}

\begin{table}[!t]
\centering
\small
\caption{Bounds for both particle-cost definitions under Theorem~\ref{thm:three-resource}, with $d=\min\{b,m,1\}$. The corresponding $\Theta$ orders are obtained when lower and upper orders match. 
}
\label{tab:resource-frontier}
\setlength{\tabcolsep}{4pt}
\begin{tabular}{@{}lccc@{}}
\toprule
Observations & Regime & Lower bound & Upper bound\\
\midrule
Interval counts & $d=0$ & $\Omega(K^2)$ & $O(K^2)$\\
& $0<d<1$ & $\Omega(K^{2-d})$ & $O(K^{2-d}\ln K)$\\
& $d=1$ & $\Omega(K\ln K)$ & $O(K\ln K)$\\
\midrule
All arrival times & $b=0$ & $\Omega(K^2)$ & $O(K^2)$\\
& $0<b<1$ & $\Omega(K^{2-b})$ & $O(K^{2-b}\ln K)$\\
& $b\ge1$ & $\Omega(K\ln K)$ & $O(K\ln K)$\\
\bottomrule
\end{tabular}
\end{table}

The group definitions give
$G(K,T_K,\JR_K)=\Theta(K^{\min\{b,m,1\}})$ and
$G_{\mathrm{full}}(K,T_K)=\Theta(K^{\min\{b,1\}})$.
Substitution in Theorem~\ref{thm:three-resource}, with $T_{\min}=c_t\tau_0$, gives Table~\ref{tab:resource-frontier}.
Since $\ln\ln K/\ln K\to0$, the limits of log optimal cost divided by $\ln K$ are $\max\{2-b,2-m,1\}$ for interval counts and $\max\{2-b,1\}$ for full records, under either cost definition.
Thus reducing the quadratic exponent requires both duration and interval-count growth in the first model, but only duration growth for full records.
Table~\ref{tab:resource-frontier} distinguishes the matching endpoints from the intermediate logarithmic gap.

Without assuming polynomial resource growth, fix $C_0>0$ and arbitrary positive $T_K$.
For either cost at most $C_0K\ln K$ and error at most $\epsilon$ for all sufficiently large $K$, interval-count observations require
\begin{equation}
\JR_K\ge c_1\frac{K}{1+\ln K},
\qquad
\frac{T_K}{\tau_0}\ge c_2\frac{K}{1+\ln K}.
\label{eq:near-linear-necessary-resources}
\end{equation}
Here $c_1,c_2>0$ depend only on $C_0,\epsilon$ and the fixed channel parameters.
The first bound follows from~\eqref{eq:expected-count-success} with fixed subset size, the second from Corollary~\ref{cor:expected-emissions}, using $F(T_K)\le p_\infty$ and $B(\mathcal C)\le\Ntot$.
Full records still require the duration bound.
Conversely, $\JR_K=\Theta(K)$ and $T_K=\Theta(\tau_0K)$ permit $\Theta(K)$ groups, so~\eqref{eq:grouped-ook-dose} achieves $O(K\ln K)$ under both costs.
The same duration suffices for full records, without an interval-count constraint.
Necessary and sufficient resource growth therefore differ by a logarithmic factor.

\section{Numerical results and discussion}
\label{sec:numerical}

\begin{figure}[!t]
\centering
\includegraphics[width=\columnwidth]{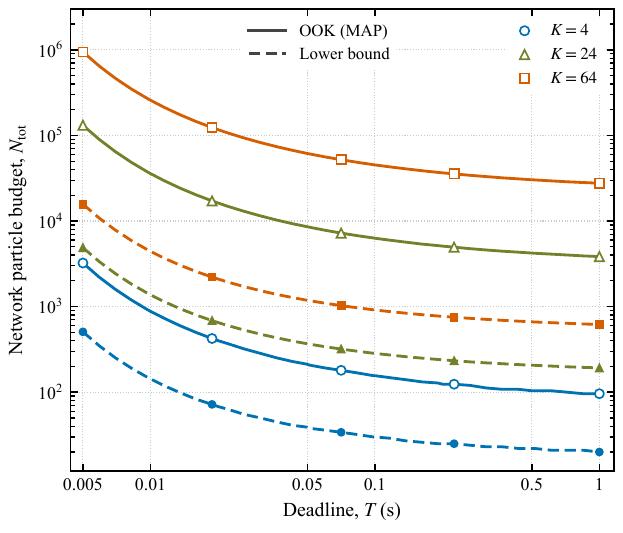}
\caption{Particle-budget bounds.}
\label{fig:actual-molecule-budget}
\end{figure}

\begin{figure*}[!t]
\centering
\includegraphics[width=\textwidth]{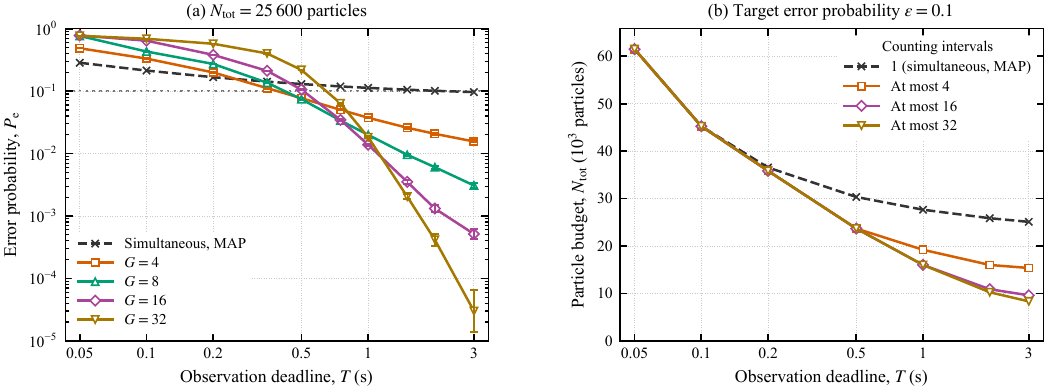}
\caption{Simultaneous OOK and GOC are compared using (a) error probability and (b) particle budget.}
\label{fig:reliability}
\end{figure*}

\subsection{Numerical setup}

For the diffusion example, take $\Dm=100\,\mu\mathrm m^2/\mathrm s$, $\aR=2\,\mu\mathrm m$, and $r_0=4\,\mu\mathrm m$, giving $\aR/r_0=1/2$ and $\tau_{\mathrm d}=0.01\,\mathrm s$.
Inputs are independent and equiprobable; each active transmitter releases $N$ particles.
At the same $K,N$, both schemes have maximum cost $\Ntot=KN$ and expected cost $KN/2$.

For simultaneous OOK, we use the full conditional binomial support with the $\Bin(K,1/2)$ prior to compute the MAP error. 
For each simultaneous-OOK budget, we employ an exhaustive search over integer $N$ to find the smallest value meeting the error target. 
We use the GOC sequential decoder described in Subsection~\ref{sec:grouped-ook} together with $\JR=G$ equal counting intervals and group spacing $\Delta=T/G$. 
We sample interval counts from the exact multinomial arrival law conditional on each group's bit sum. 
We retain earlier-group delayed arrivals, whereas arrivals after $T$ and particles that never arrive are unobserved. 

\begin{figure*}[!t]
\centering
\includegraphics[width=\textwidth]{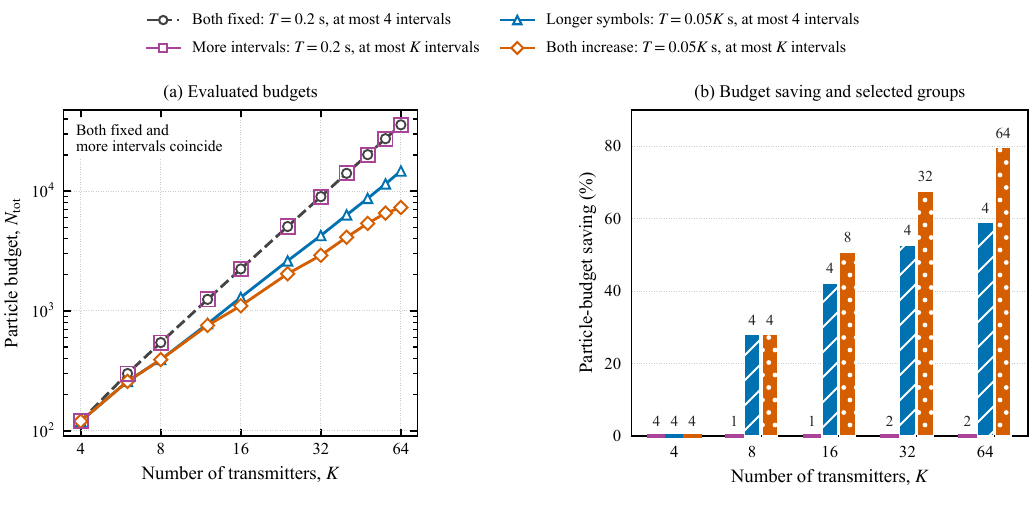}
\caption{Particle budgets and selected groups under four resource conditions.}
\label{fig:network-scaling}
\end{figure*}

For GOC, we use $200{,}000$ Monte Carlo trials per point in Fig.~\ref{fig:reliability}(a) and per candidate confirmation in Figs.~\ref{fig:reliability}(b) and~\ref{fig:network-scaling}.
Each GOC budget candidate is fixed before its own confirmation, which uses fresh trials independent of the search samples.
For each budget figure separately, we use Bonferroni-corrected one-sided exact binomial upper confidence limits at significance level $0.05/m$, where $m$ is the stated maximum family size.
The resulting limits have at least $95\%$ joint coverage within that figure.
For budget selection, we retain only GOC candidates whose upper limits are at most $0.1$.

\subsection{Particle budget and observation deadline}

Figure~\ref{fig:actual-molecule-budget} compares finite lower bounds with simultaneous-OOK budgets for $K=4,24,64$, $\Pe\le0.1$, and one counting interval. 
We obtain the lower curves by inverting~\eqref{eq:r15-count-binary}, \eqref{eq:subset}, and~\eqref{eq:simple-nonarrival}, rounding each resulting budget up to an integer, and taking the largest value. 
The nonarrival bound determines all plotted lower-bound points. 

For simultaneous OOK at fixed $N$, increasing $T$ raises the absorption probability, which makes conditional count distributions for neighboring sums easier to distinguish. 
For $K=64$, the simultaneous-OOK budgets for $T$ equal to $0.01$ and $0.05\,\mathrm s$ are $257{,}856$ and $61{,}504$ particles, respectively. 
Extending $T$ from $0.5$ to $1\,\mathrm s$ reduces the budget by only about $9\%$. 
The reason for the diminishing benefit is that $F(T)$ approaches the eventual absorption probability $\aR/r_0=1/2$. 

At $K=64$ and $T=1\,\mathrm s$, however, the simultaneous-OOK budget remains about $45$ times the plotted lower bound. 

\subsection{Reliability at a fixed particle budget}

Figure~\ref{fig:reliability}(a) shows error probability versus $T$ for the schemes with $K=64$, $N=400$, and $\Ntot=25{,}600$. 
The bars indicate pointwise $95\%$ Wilson confidence intervals for GOC.
At $T=0.2\,\mathrm s$, the error is about $0.167$ for simultaneous OOK and $0.382$ for $G=16$, respectively. 
At $T=1\,\mathrm s$, the estimated error for $G=16$ is much lower at $0.0139$, about $88\%$ below the simultaneous-OOK error. 

The group size and hence the conditional variance bound in Subsection~\ref{sec:grouped-ook} decrease as $G$ increases. 
At fixed $T$, we observe that as $\Delta=T/G$ decreases, $h_0$ decreases and the separation between the decoder's group-count means decreases. 
Since the sequential decoder subtracts estimated earlier-group mean contributions, delayed-arrival fluctuations remain and earlier decision errors can affect later estimates. 
The curve crossings show that choices among these complete transmission and decoding schemes depend on symbol duration. 

At $K=64$, $N=400$, and $T=1\,\mathrm{s}$, the mean total absorbed count is about $5680$ for simultaneous OOK and $5217$ for GOC with $G=16$. 
Later releases leave less time for absorption, reducing the mean by about $8.2\%$ at the same release budget. 
For simultaneous OOK, the total absorbed count is sufficient for the input sum. 
More counting intervals within the same symbol duration therefore cannot reduce the optimal error probability. 
The lower error here is consistent with the joint use of staggered releases and counting intervals. 

\subsection{Particle budget with limited counting intervals}

Figure~\ref{fig:reliability}(b) shows the achievable budgets from the evaluated schemes for $K=64$ and $\Pe\le0.1$ under counting-interval limits of $1$, $4$, $16$, and $32$. 
We evaluate both simultaneous OOK and GOC with $G\in\{2,4,8,16,32\}$, using an $N$ grid with $10$-particle steps for GOC. 

The correction uses a maximum family of $m=35$ possible GOC candidates, covering all combinations of seven deadlines and five group counts.
Within each interval limit, we select the smallest evaluated budget, preferring fewer groups in ties.

For every curve, simultaneous OOK is selected at $T=0.05$ and $0.1\,\mathrm s$. 
For interval limits of $4$, $16$, or $32$, $G=2$ is selected for all three curves at $T=0.2\,\mathrm s$ and $G=4$ at $T=0.5\,\mathrm s$. 
At $T=1\,\mathrm s$, we observe that interval limits of $16$ or $32$ both select $G=16$, which requires $16{,}000$ particles, about $42\%$ below the simultaneous-OOK budget. 
Achieving the lowest confirmed budget among evaluated candidates therefore does not require all available counting intervals. 
At $T=3\,\mathrm s$, $32$ groups reduce the budget by about $67\%$ compared with simultaneous OOK. 

\subsection{Network size and resource growth}
\label{sec:network-scaling}

Figure~\ref{fig:network-scaling}(a) compares particle budgets for $\Pe\le0.1$ at eleven network sizes from $4$ to $64$. 
For each condition, we select the smallest evaluated budget among the exact simultaneous OOK baseline and confirmed GOC candidates. 
The GOC candidates use $G=2,4,8,\ldots$ not exceeding either $K$ or the counting-interval limit. 
Group sizes differ by at most one transmitter, with larger groups releasing first. 

The correction uses a maximum family of $m=86$ possible GOC candidates across the conditions in Fig.~\ref{fig:network-scaling}.

At $T=0.2\,\mathrm s$, increasing the interval limit from $4$ to $K$ does not reduce the selected budget within this candidate family. 
At $K=64$, joint growth selects $G=64$ and $7296$ particles. 
The selected budgets are $35{,}776$ when both resources are fixed and $14{,}720$ with $G=4$ when only symbol duration increases. 
Joint growth therefore saves $79.6\%$ and $50.4\%$, respectively. 
Joint growth allows more groups while keeping the selected counting intervals at least $0.05\,\mathrm s$ long. 

These finite comparisons illustrate the resource tradeoff proved in Subsection~\ref{sec:resource-exponents} for interval-count observations. 
A smaller budget exponent requires both symbol duration and the number of counting intervals to grow. 

\section{Conclusion}
\label{sec:conclusion}

For independent equiprobable bits under the stated arrival conditions, the minimum budget for exact counting is quadratic in network size at a fixed symbol duration and error target, even when all arrival times are observed. 
At fixed symbol duration, any subquadratic budget makes the error probability approach one for every allowed release rule and decoder. 
These results also hold with only the expected number of released particles constrained. 
The lower bounds and GOC determine the optimal polynomial budget exponents when the symbol duration and, for interval-count observations, the number of intervals are fixed or grow polynomially with network size. 

For the diffusion example, the numerical results show that GOC provides achievable budget savings and demonstrate that the selected group count depends on the symbol duration. 
The finite-network optimum, the leading constants in the scaling laws, and the intermediate logarithmic gaps in Subsection~\ref{sec:resource-exponents} remain open problems. 

\end{document}